\documentclass[prl,aps,twocolumn,showpacs]{revtex4}
\usepackage{psfrag}
\usepackage{graphicx}
\usepackage{graphics}
\usepackage{bm}
\usepackage{color}

\newcommand{\be}{\begin{equation}}
\newcommand{\ee}{\end{equation}}
\newcommand{\ba}{\begin{eqnarray}}
\newcommand{\ea}{\end{eqnarray}}

\newcommand{\dcom}[1]{}
\newcommand{\dnote}[1]{}

\newcommand{\gsim}{\raise.3ex\hbox{$>$\kern-.75em\lower1ex\hbox{$\sim$}}}
\newcommand{\lsim}{\raise.3ex\hbox{$<$\kern-.75em\lower1ex\hbox{$\sim$}}}

\begin{document}

\renewcommand{\thefootnote}{\fnsymbol{footnote}}

%-------------------------------------------------

\renewcommand{\thefootnote}{\arabic{footnote}}
\setcounter{footnote}{0} \typeout{--- Main Text Start ---}

\title
{The Role of Causality in Tunable Fermi Gas Condensates}
\author{ Jen-Tsung Hsiang, Chi-Yong  Lin, and Da-Shin  Lee}\affiliation{
Department of Physics, National Dong Hwa University, Hua-Lien,
Taiwan 974, R.O.C.  }
\author{Ray\ J.\ Rivers}
\affiliation{ Blackett Laboratory, Imperial College\\
London SW7 2BZ, U.K.}

\date{\today}
\begin{abstract}
We develop a new formalism for the description of the condensates of  cold Fermi atoms whose speed of sound can be tuned with the aid of a narrow Feshbach resonance.  We use this to look for spontaneous phonon creation that mimics spontaneous
particle creation in curved space-time in Friedmann-Robertson-Walker
and other model universes.
\end{abstract}


\maketitle
\section{Introduction}
There has been a very fruitful exchange of ideas between particle/astro-physicists and condensed matter theorists over several decades.  Specifically, black holes and other
phenomena require an improved understanding of semiclassical gravity
and the role of the even more inaccessible Planck scale. Causality
plays a key role in this and, given that direct
tests are experimentally impossible, analogies have been sought in
condensed matter physics that seek to replicate this causality in
the laboratory.
Analogies with black holes are direct due to the pioneering work of Unruh\cite{unruh},
which showed the similarities between {\it event horizons} in fluids (most
simply when flow rate exceeds the local speed of sound) and the event horizons of black holes. The
programme of 'analogue gravity' begun to exploit these similarities  has generated over
700 papers since its implementation [see the recent review article \cite{visser}].

One of the great achievements of the last several years has been the
construction of condensates in which the speed of sound $c_s$ is
tunable. Systems whose causal behaviour is so simply controlled are potentially good candidates with which to explore analogies.
In this paper we explore the way in which tunable
gases of cold Fermionic atoms, which pair to form bosonic condensates,  can
produce and propagate phonons in mimickry of particle production in the early universe.
Before doing so we need to address a basic problem in analogue gravity modelling. This is that there is an immediate mismatch; the early
(and late) universe is relativistic, usually locally Lorentzian,
whereas condensed matter systems and condensates are
Galilean in their space-time symmetry.  In practice we
rely on the gapless phonon mode (with dispersion relation $\omega =
c_s k$) to mimic the relativistic behaviour of light in the geodesic
equations that control its behaviour.
However, at best the parallel breaks down at the Debye scale
(the minimum  wavelength of the phonon mode). This breakdown of speed-of-sound (SoS) Lorentzian behaviour has been
considered \cite{visser2,visser3} a counterpart to the possible breakdown of
true Lorentzian behaviour at the Planck scale.

For the purpose of drawing analogies with phonon production, and more generally, it is important to develop a formalism that interpolates between the SoS Lorentzian behaviour of the phonon, where appropriate, and the underlying Galilean invariance that provides the bedrock to all our analysis. The main result of this paper is to provide such a formalism for Fermi gases whose speed of sound is tunable in an external magnetic field by virtue of a {\it narrow} Feshbach resonance.

The organisation of the paper is as follows. We first show how the speed of sound in di-fermion (or diatomic)  condensates is tunable by the application of an external magnetic
field. This is because the effective bosonic theory is not the phonon Goldstone model of
the familiar Gross-Pitaevskii (GP) equation, but a Higgs-Goldstone model
where the Higgs field is represented by the gapped diatom density
fluctuations. We have discussed this
elsewhere \cite{rivers,rivers2,rivers3,rivers4}, to which the reader
is referred for greater detail. Here we take a
further step by tracing out the density fluctuations to give
an effective low-energy phonon model for the phonon field $\theta$ alone. We then use this to discuss spontaneous phonon production in a rapid magnetic field quench.

There is a problem of familiarity (or lack of it) with Galilean invariance. We work with a Lagrangian formalism rather than the more familiar canonical Hamiltonian approach. Inevitably, we have to make approximations and it is crucial that these approximations preserve the underlying Galilean symmetry. Our results are straightforward to those readers familiar with explicit Galilean invariant approaches to condensed matter theory as in \cite{greiter,aitchison,son}, for example. For those readers less familiar, sufficient to say that the fundamental Galilean phonon scalar is the combination
\be
G(\theta) = \dot{\theta} + (\nabla
\theta )^2/2M
\ee
for di-atoms with mass $M=2m$, where $m$ is the atomic mass.

Our main result is that the phonon action takes the non-local form
\begin{eqnarray}
&& S_{eff}[\theta] = \int d^4x \left\{\frac{N_0}{4}\ G^2(\theta (x)) -\frac{1}{2}{\rho}_0
   G(\theta (x) )    \right\}
 \nonumber\\
 && +\frac{\alpha^2}{2}\int d^4x \int d^4x' G(\theta
 (x))\, D_F (x, x') \, G(\theta
 (x')) \, , \nonumber\\
 \label{Seff_thetaa}
 \end{eqnarray}
with coefficients to be discussed later, where $i D_F (x, x')$ describes the exchange of virtual density fluctuations in the condensate as diatoms associate and disassociate (with constant overall density). This density field can be understood as the Galilean scalar 'Higgs' boson (mentioned earlier) to complement the Goldstone phonon. Its properties, as well as its coupling strength $\alpha$ depend on the external magnetic field with which we tune the binding energy of the Feshbach resonance.
The form (\ref{Seff_thetaa}) is a straightforward generalisation of the Galilean invariant action of \cite{aitchison} and a natural relation to that of \cite{son}, for example.

The SoS Lorentzian limit is obtained from the Galilean formulation by the substitution \cite{greiter,aitchison}
\be
G(\theta)\rightarrow (\nabla\theta )^2/2M,\,\,\,\,\,\,\,G^2(\theta)\rightarrow\dot{\theta}^2
\ee
The outcome is the action
\begin{eqnarray}
S_{eff}^{qu}[\theta] &=& \frac{1}{4}\int d^4x  \{N_0{\dot\theta}^2(x) -\frac{\rho_0}{2m}
   (\nabla\theta)^2\}
   \nonumber
   \\
&& +\frac{\alpha^2}{2}\int d^4x \int d^4x'{\dot\theta}
(x)\, D_F (x-x') \, {\dot\theta}(x').
\label{Squad_thetab}
 \end{eqnarray}
The details will be given later but it is already clear that this action essentially describes an SoS Lorentzian phonon, allowing for some 'fuzzyness' in the sound-cone due to density fluctuations.

To bring this back to familiar territory, in those regimes for which the action (\ref{Seff_thetaa}) is approximately local with the fuzzyness smoothed away, the resulting equations of motion can be rewritten in the Gross-Pitaevskii form
 \be
 i\hbar\dot{\psi} + \frac{\hbar^2}{2m}\nabla^2\psi + 2mc_s^2\psi - \frac{2mc_s^2}{\rho_0}\psi|\psi|^2 = 0,
  \label{GP}
 \ee
where $\psi = \sqrt{\rho}\;\exp(i\theta)$ for a condensate of
density $\rho$, with $\theta$ the phonon field with speed of sound $c_s$. We have used this GP equation elsewhere \cite{rivers3} to examine possible spontaneous vortex production in field quenches  but we shall not pursue that here  and work from (\ref{Seff_thetaa}).

The action (\ref{Seff_thetaa}) permits different approximations. A phonon moving with variable speed $c_s$ in space and time can be interpreted as moving along a geodesic path in a varying space-time metric. This is the basis of analogue gravity, for which the geodesic equation for the phonon following from (\ref{Seff_thetaa}) in the hydrodynamic limit is
\be
 \frac{1}{\sqrt{-g}}\partial_{\mu}(\sqrt{-g}~g^{\mu\nu}\partial_{\nu}\theta ) = 0,
 \label{geodesic}
\ee
where $g$ is the acoustic (or hydrodynamic) metric of the condensate. For example, for homogeneous
condensates it is relatively simple experimentally to quench the condensate so that its
acoustic metric takes Freedman-Robinson-Walker (FRW) form. Spontaneous phonon production in this metric is the counterpart to
the familiar process of particle production in the early FRW
universe and it is this analogy that we pursue, to demonstrate how simple the formalism is. Despite the
underlying Galilean invariance, it has been argued
\cite{jain1,jain2,jain3} that there is still a window in which
geodesic behaviour of (\ref{geodesic}) dominates.
We will examine this below.

We stress that for us the default is the underlying Galilean invariance for which the SoS Lorentzian limit may sometimes be appropriate.

\section{Condensates from cold Fermi gases}
We now show how a cold Fermi gas can lead to a condensate described by the action (\ref{Seff_thetaa}). The forces between alkali fermi atoms are such that, for weak
coupling, they form Cooper pairs (the BCS regime), correlated in
momenta whereas, for strong coupling they form molecules or diatoms
(the BEC regime), correlated in position. In a two-channel model
with a narrow Feshbach resonance the binding energy of this
resonance can be changed by the application of an external magnetic
field because of the different Zeeman effects between the channels.
The outcome is to change the strength of the force and permit us to
take the system smoothly from the BCS regime (with negative s-wave
scattering length $a_S$) to the BEC regime (with positive $a_S$)
through the {\it unitary limit} at which $a_S$ diverges. The speed
of sound also varies smoothly through this crossover  from $c_{BCS}
= v_F/\sqrt{3}$ in the deep BCS regime, where $v_F$ is the Fermi
velocity, to $c_s\rightarrow 0$ in the deep BEC regime. This
remarkable control over the sound speed by the simple application of
an external field gives us control over the causal properties of the
gas.

Analytically this is most simply understood if we take the resonance to be very
narrow, for which the mean-field
equation, necessary for analytic approximations, can be trusted \cite{gurarie}.
Further, for a narrow resonance the condensate order parameter is the Feshbach resonance field itself which, as we have shown \cite{rivers4}, leads to a single-fluid model in the hydrodynamic limit.

Our starting point
is the exemplary 'two-channel' microscopic action\cite{timmermans,gurarie} (in units in which
$\hbar = 1$)
\begin{eqnarray}
S &=& \int dt\,d^3x\bigg\{\sum_{\uparrow , \downarrow}
 \psi^*_{\sigma} (x)\ \left[ i \
\partial_t + \frac{\nabla^2}{2m} + \mu \right] \ \psi_{\sigma} (x)
\nonumber \\
   &+& \varphi^{*}(x) \ \left[ i  \ \partial_t + \frac{\nabla^2}{2M} + 2 \mu -
\nu \right] \ \varphi(x) \nonumber \\
&-& g \left[ \varphi^{*}(x) \ \psi_{\downarrow} (x) \ \psi_{\uparrow}
(x) + \varphi(x)  \psi^{*}_{\uparrow} (x) \ \psi^{*}_{\downarrow} (x)
\right]\bigg\} \label{Lin}
\end{eqnarray}
for cold ($T = 0$) fermi fields $\psi_{\sigma}$
 with spin label $\sigma = (\uparrow, \downarrow)$.
 The diatomic field $\varphi$ describes a {\it narrow} bound-state (Feshbach) resonance with
 tunable binding energy $\nu$ and mass $M =2m$.
 The idealisation of ignoring the self-interactions between fermions and  diatomic molecules is well suited for describing the crossover phenomena, but breaks down in the deep BCS and deep BEC regimes \cite{gurarie}, although not in such a way as to compromise our results.

 We restrict ourselves to the {\it mean-field
approximation}. ${S}$ is quadratic in the fermi fields. Integrating them out gives an efective action for the order parameter  $ \varphi (x) = -|\varphi(x)| \ e^{i
\theta (x)}$ alone.
This action
possesses a $U(1)$ invariance under
$\theta\rightarrow\theta +\rm{const.}$,
 which
is spontaneously broken by spacetime constant {\it gap} solutions $|\phi (x)| = |\phi_0|\neq
 0$. We expand in fluctuations about $\phi_0$, but {\it not} with the decomposition $\phi = \phi_0 + \delta\phi$, since we need to preserve Galilean invariance at each step of the approximation.

The Galilean invariants of the theory are the
density fluctuation
 $\delta |\phi|= |\phi|  -  |\phi_{0}|$,
$G(\theta) = \dot{\theta} + (\nabla
\theta )^2/4m$ as mentioned earlier, and  $D_t(\delta |\phi|,\theta ) =
\dot{(\delta |\phi|)}+ \nabla \theta .\nabla (\delta |\phi|)/2m$. $D_t$ is the comoving time derivative of $\delta |\phi|$ in a fluid with fluid velocity $\nabla \theta/2m$. $\theta (x)$ is not small. Let us rescale $\delta|\phi|$ to $\delta|\phi|=\kappa\epsilon$. The resulting Galilean invariant effective
action for the long-wavelength, low-frequency condensate takes the form \cite{rivers,rivers2}
      \begin{eqnarray}
 S_{eff}[\theta, \epsilon] &=& \int d^4x\bigg[ \frac{N_0}{4}\ G^2(\theta, {\epsilon}) -\frac{1}{2}{\rho}_0
   G(\theta, {\epsilon})
 \nonumber\\
 && -{\alpha}{\epsilon}G(\theta, {\epsilon})
 +\frac{1}{4}{\eta}D_t^2({\epsilon},\theta)
  -\frac{1}{4}{\bar M}^2{\epsilon}^2\bigg].
 \label{LeffU0}
 \end{eqnarray}
The scale factor $\kappa$, defined in the Appendix,
%scaling
is chosen so that on extending
$G(\theta)$ to $G(\theta, \epsilon ) = \dot{\theta} + (\nabla
\theta )^2/4m + (\nabla \epsilon )^2/4m$, $\epsilon$ has the
 same coefficients as $\theta$ in its spatial derivatives.

The quadratic part $S_{eff}^{qu}$ of $S_{eff}$ is
       \begin{eqnarray}
 S_{eff}^{qu}[\theta, \epsilon] &=& \int d^4x\bigg[\frac{N_0}{4}{\dot\theta}^2 -\frac{{\rho}_0}{8m}(\nabla
\theta )^2
 \nonumber\\
 && -{\alpha}{\epsilon}\dot\theta
 +\frac{1}{4}{\eta}{\dot\epsilon}^2
  -\frac{{\rho}_0}{8m}(\nabla
\epsilon )^2
  -\frac{1}{4}{\bar M}^2{\epsilon}^2\bigg].
 \label{Leff2}
 \end{eqnarray}
We see immediately from (\ref{Leff2}) that the action describes a gapless (i.e. SoS Lorentzian, or SoS relativistic) Goldstone mode $\theta$, the phonon, and a gapped (but also relativistic) Higgs mode. The underlying Galilean invariance is only present in their time-derivative coupling with strength $\alpha$.

As for the coefficients in (\ref{Leff2}),
  $\rho_0 = \rho^F_0 +
\rho^B_0 $ is
 the total (fixed) fermion number density where
$\rho^F_0$  is the explicit fermion density
  and $\rho^B_0 = 2|\phi_{0}|^2$ is
due to molecules (two fermions per molecule).
For the evolving system the molecular or diatomic density is
$\rho^B = 2|\phi|^2 = \rho^B_0 + 4\delta |\phi||\phi_0|$. This shows that $\epsilon\propto \delta\rho^B$ is the scaled molecular density fluctuation, describing the repeated dissociation of molecules into atom pairs and their reconversion into molecules \cite{timmermans,gurarie2}.
Otherwise, the coefficients are somewhat opaque and relegated to the Appendix.
We give exemplary plots in Fig.1 of the most important combinations.
\begin{figure}
\centering
\includegraphics[width=\columnwidth]{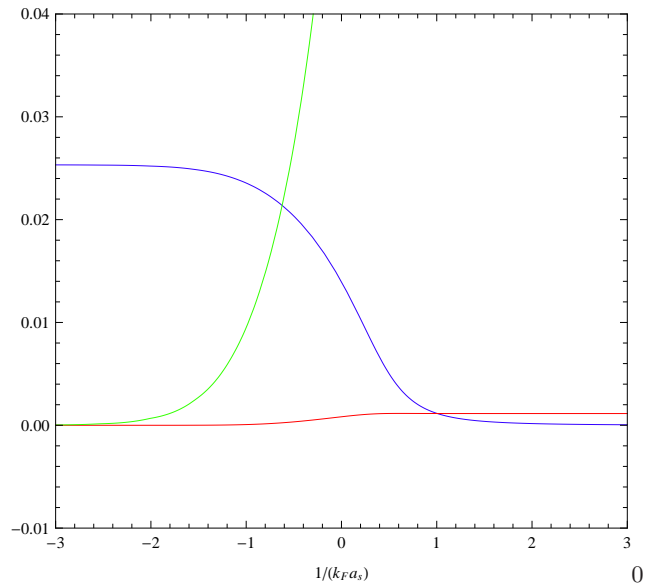}0
\caption{
The curves show ${\alpha^2}$ (red), $N_0$ (blue) and ${\alpha^2}/{\bar M}^2$ (green) for the value ${\bar g} = 0.9$ (defined later in terms of $g$) as a function of $1/k_Fa_S$.
%\textcolor{red}
At large values of $1/k_Fa_S$ $\alpha\approx \rho_0$, as chosen later.}
\end{figure}

We observe that $S_{eff}[\theta, \epsilon]$ of (\ref{LeffU0}) is quadratic in $\epsilon$. Integrating out the $\epsilon$ field gives the effective non-local action of (\ref{Seff_thetaa}) in terms of the phonon alone,
\begin{eqnarray}
&& S_{eff}[\theta] = \int d^4x \left\{\frac{N_0}{4}\ G^2(\theta (x)) -\frac{1}{2}{\rho}_0
   G(\theta (x) )    \right\}
 \nonumber\\
 && +\frac{1}{2}\int d^4x \int d^4x' {\alpha}(x) G(\theta
 (x))\, D_F (x, x') \, {\alpha (x') } G(\theta
 (x')) \, , \nonumber\\
 \label{Seff_theta}
 \end{eqnarray}
where $i D_F (x, x')=\langle T(\epsilon (x)
\epsilon(x'))\rangle_0$, determined from the quadratic action $S_{eff}[0,\epsilon]$ of (\ref{Leff2}) when $\theta\equiv 0$, and we now allow $\alpha$ to be space-time dependent.

As we said in the introduction, this is the main formal result of the paper, showing intuitively how virtual (Higgs) density fluctuations mediate the phonon field in an explicitly Galilean invariant way.
For constant parameters the quadratic part of $S_{eff}[\theta]$ was already given in (\ref{Squad_thetab}) as
\begin{eqnarray}
 S_{eff}^{qu}[\theta] &=& \frac{1}{4}\int d^4x  \{N_0{\dot\theta}^2(x) -\frac{\rho_0}{2m}
   (\nabla\theta)^2\}
   \nonumber
   \\
&& +\frac{\alpha^2}{2}\int d^4x \int d^4x'{\dot\theta}
(x)\, D_F (x-x') \, {\dot\theta}(x')
\label{Squad_theta}
 \end{eqnarray}
but with the further information that
\begin{eqnarray}
D_F (x - x') = -\frac{2}{(2\pi)^4}\int d\omega~d^3 k\frac{e^{i\omega(t-t')}e^{-i{\bf k}{\bf(x - x')}}}{\eta\omega^2 - \rho_0k^2/2m - {\bar M}^2}.
%\nonumber\\
\label{corr}
\end{eqnarray}
To understand the fuzzyness of the sound cone due to diatom density fluctuations we observe that,
in the BEC regime, for which $\eta\approx 0$, we can ignore the $\eta\omega^2$ term in the denominator of the propagator for all but the shortest wavelengths. The resulting propagator is local in time, making the sound-cone smooth. However, as we move into the BCS regime both $\eta$ and ${\bar M}^2$ increase and the two time derivatives in the second term of (\ref{Squad_theta}) are split. There is compensation in that $\alpha$ becomes smaller, vanishing in the deep BCS regime, but there is a characteristic frequency of density oscillations associated with the non-local smearing of the soundcone. It is mainly in the BEC regime that the speed of sound changes sufficiently rapidly for causal effects to be important. What the above shows, and which we have shown elsewhere by different means \cite{rivers3}, is that we cannot push the system much out of the BEC regime, if at all, before analogies break down because the oscillations are too slow to be ignored \cite{gurarie2,timmermans}.

From Eq.(\ref{Squad_theta}) we read off the dispersion relation:
 \begin{eqnarray}
 \omega^2\bigg(N_0 + \frac{4\alpha^2}{-\eta\omega^2 + \rho_0k^2/2m + {\bar M}^2}\bigg) - \frac{\rho_0}{2m}{\bf k}^2 = 0,
 \label{disperse}
 \end{eqnarray}
a result not achieved so transparently before.
In general  we allow all parameters to vary in
space-time, although subsequently we restrict ourselves to
homogeneous systems.

 \section{The hydrodynamic (or acoustic) approximation}

The hydrodynamic (or acoustic) approximation is obtained by approximating  the  SoS relativistic Higgs propagator $ D_F (x, x') $ by its long wavelength, low energy limit, the instantaneous contact term
\begin{equation}
D_F (x- x') \approx \frac{2}{{\bar M}^2} \delta (x-x') \, ,
\label{feynprop_hydro}
\end{equation}
(whether ${\bar M}^2$ is constant or not). In this ultra-local approximation the density fluctuations do not propagate. That is, in terms of the two-field action (\ref{Leff2}) we are neglecting the spatial and temporal variation  of $\epsilon$, in comparison to $\epsilon$ itself, equivalent to setting $\omega$ and $\bf k$ to zero in the denominators of (\ref{corr}) and (\ref{disperse}).

This corresponds to making the approximation
\begin{eqnarray}
S_{hyd}[\theta]
&=&
\int d^4x\bigg[\frac{1}{4}(N_0 + 4\alpha^2/{\bar M}^2)\ G^2(\theta (x))
\nonumber\\
&-&
\frac{1}{2}{\rho}_0
   G(\theta (x) )\bigg]
 \label{Seff_theta0}
 \end{eqnarray}
for the action (\ref{Seff_theta}) in which we have taken the coefficients to be constant.  In the BCS regime, when $\alpha \approx 0$ this is just the form of the action given in \cite{aitchison} in the same approximation. [However, in this latter case there was no explicit resonance.]
  It is also the (exact) long wavelength limit of (\ref{disperse}), in which the phonon has the linear dispersion relation
 $\omega^2 = c_s^2 {\bf k}^2$,
 with speed of sound
  \be
  c_s^2 = \frac{\rho_0/2m}{N_0 + 4{\alpha}^2/{\bar M}^2},
  \label{v-s}
 \ee
 as can be read directly from (\ref{Seff_theta0}) on expanding $G$.
 In the deep BCS regime $c_s\rightarrow c_{BCS} = \sqrt{\rho_0/2m N_0} = v_F/\sqrt{3}$. On the other hand, in the deep BEC
regime (where $N_0$ and ${\bar M}^2$ are small, and $\alpha = \rho_0$), then $c_s\rightarrow 0$. To justify the acoustic/hydrodynamic terminology, we note that the equation of motion following from (\ref{Seff_theta0}) can be written as a continuity equation in which the density fluctuations satisfy the Bernoulli equation \cite{aitchison}.

An exemplary graph of $c_s^2$ is given in Fig.2.
\begin{figure}
\includegraphics[width=\columnwidth]{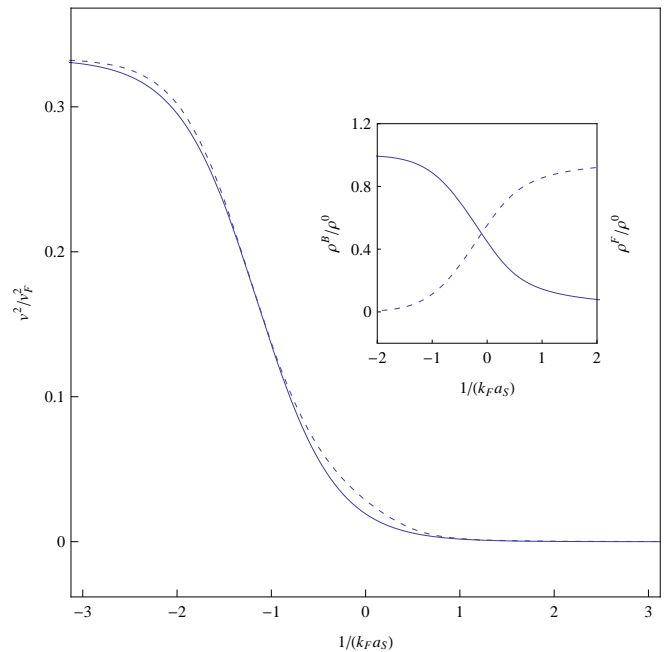}
\caption{
The dotted line shows $c^2$ for the value ${\bar g} = 0.9$ as a function of $1/k_Fa_S$. The solid line shows the parametrization (\ref{tanh}) for $d(\bar g) = -1.4$ and $b(\bar g) = 1.2$. We get as good or better fits for other values of $\bar g$, with $b(\bar g)$ varying by only 25\% over the range $0.2 \leq {\bar g}\leq 1.6$.}
\end{figure}
To a fair approximation, $c_s^2$ can be approximated as
\be
 c_s^2\approx (c_{BCS}^2/2)[1+ \tanh (d(g) - b(g)/k_Fa_S)].
 \label{tanh}
\ee
%\textcolor{red}{}
This has the great virtue that calculations can be performed
analytically. Empirically, $b(g)$ is insensitive to the
coupling constant. In fitting $d(g)$ it happens that,
as we reduce $g$ the best-fit to $d(g)$ becomes more negative. Further, on going from
the BCS toward the BEC regime, for a smaller value of  $g$ the sound
speed changes earlier and more quickly.

We conclude with an observation on the importance for the calculations above of the Feshbach resonance being narrow. If the resonance is not narrow direct atomic interactions occur, manifest through a four-fermi interaction in (\ref{Lin}). The order parameter now has two terms, a resonance contribution and a di-fermion field characterizing this interaction \cite{ohashi,rivers}. In the same acoustic/hydrodynamic limit as above the equations of motion can now be understood, in general, as representing two coupled fluids \cite{rivers,rivers2}.  Only when the contact terms overwhelm the resonance exchange do we again recover a single fluid model \cite{aitchison}(in the hydrodynamic approximation). Otherwise the single GP equation (\ref{GP}) is replaced by two coupled GP equations \cite{rivers2} or the geodesic equation (\ref{geodesic}) has two metrics, neither of which lends itself to simple analogy.

\subsection{Tuning the condensate}

For our narrow resonance, applying a {\it homogeneous} external magnetic field ${\cal H}$ changes $c_s$ by changing the s-wave scattering length $a_S$ as\cite{gurarie}
 \be
 \frac{1}{k_F a_S} \propto({\cal H} - {\cal H}_0),
 \label{Hdep}
\ee
 where  ${\cal H}_0$ is the field required to achieve the unitary limit ($|a_S|\rightarrow\infty$). For our first case of interest we pass from the BEC to the BCS regimes as ${\cal H}$ {\it increases} through ${\cal H}_0$.

If we adopt (\ref{tanh}) the resulting equations can then be solved analytically, a good approximation \cite{rivers3} even though the $tanh$-behaviour slightly overestimates the rate at which the speed of sound vanishes.

 In the first instance we consider simple quenches in which ${\cal H}$ increases {\it uniformly} in time for which, on going from the BCS to BES regimes we take ${\dot {\cal H}}/{\cal H}|_{{\cal H}_0} = -{\tau_H}^{-1}$.   The time dependence of $c_s(t)$ as ${\cal H}$ changes can now be written as
 \be
 c_s^2(t) = (c_{BCS}^2/2)[1 + \tanh (d(g) - t/\tau_Q)],
\ee
where $\tau_Q \propto \tau_H$.  $t=0$ is the time at which the system is at the unitary limit (i.e $t <0$ in the BCS regime, $t > 0$ in the BEC regime).
 If going from the BEC to BCS regimes we reverse $t\rightarrow -t$.

To be concrete, we consider the narrow resonance in $^6Li$ at ${\cal H}_0 = 543.25 G$, discussed in some detail in \cite{strecker}.  As our benchmark we take the achievable number density $\rho_0\approx  3 \times 10^{12} cm^{-3}$, whence $\epsilon_F\approx  7 \times 10^{-11} eV$ and $\gamma_0\approx 0.2$. In terms of the dimensionless coupling $\bar g$,
 where $g^2 = (64\epsilon^2_F/3 k_F^3){\bar g}^2$, $^6Li$ at the density above corresponds to ${\bar g}^2\lesssim 1$.

 If $\tau_0 = \hbar/\epsilon_F$, the inverse Fermi energy (in units of $\hbar$) then
 \be
 \frac{\tau_Q}{\tau_0}\approx \frac{1}{{\dot {\cal H}}},
 \ee
 where $\dot {\cal H}$ is measured in units of $Gauss\, (ms)^{-1}$.  Experimentally, it is possible to achieve quench rates as fast as ${\dot {\cal H}}\approx 0.1 G/ms$~\cite{strecker}. As we shall see below, even they are not sufficiently fast to shake off Galilean invariance.

\section{Analogue Gravity}
Rather than
consider event horizons we continue with the relatively simple case
of a homogeneous external field, looking for analogies
with Friedmann-Robertson-Walker (FRW) universes in which particle
creation can be observed, proposed in \cite{jain1,jain2,jain3} in
particular. Specifically, the analysis in \cite{jain1,jain2,jain3}
is predicated on the representation of the condensate by a
Gross-Pitaevskii (GP) mean bosonic field in which the strength of
the self-coupling, and hence the scattering length, is given an
explicit time-dependence.

We have already indicated how, for cold Fermi gases, the
{\it linear} behaviour of the long-wavelength condensate dispersion
relation can be derived from a GP equation. We have also anticipated short-wavelength non-linear behaviour, crucial for
understanding the applicability of (\ref{geodesic}), that is quantitatively similar to that shown in \cite{jain1,jain2,jain3}. However, the difference that lies in the detail is sufficient to make what seemed a difficult task in their case an impossible task here.

We stress that, while familiarity makes analogies simpler, there is nothing
intrinsically important about mimicking FRW universes and we begin
more generally, with an inhomogeneous background.

\subsection{The acoustic metric}

On allowing for explicit time dependence in the Lagrangian (\ref{Seff_theta0})
the Euler-Lagrange equation now takes the form
\be
\frac{d}{dt}\bigg[\frac{\rho_0}{c_s^2}G(\theta)\bigg] + \nabla.\bigg[\frac{\rho_0}{2mc_s^2}(G(\theta)\nabla\theta)\bigg] - \rho_0\nabla^2\theta = 0
\label{ELac}
\ee
in which $c_s$ is the local speed of sound
 as before, but now where $\rho_0, N_0$, etc. vary in space and time.

With zero background
velocity (\ref{ELac}) can be linearised as the geodesic equation
\be
 \frac{1}{\sqrt{-g}}\partial_{\mu}(\sqrt{-g}~g^{\mu\nu}\partial_{\nu}\tilde\theta ) = 0,
 \label{geodesic2}
\ee
 with which we began, where in $d$ spatial dimensions
\begin{eqnarray}
 {g}_{\mu\nu} &=& \bigg(\frac{\rho_0}{c_s}\bigg)^{\frac{2}{d-1}}\left( \begin{array}{cc}
        - c_s^2         & 0 \\
                  0 & \delta_{ij} \, .
                  \end{array} \right)
                  \label{FRWv}
                  \end{eqnarray}
\subsection{Mimicking spatially homogeneous (and FRW) universes}

In this we follow the approach of \cite{jain1,jain2,jain3}. Consider a homogeneous condensate in a homogeneous magnetic field ${\cal H}(t)$ varying in time. The phonon field $\theta$  then satisfies (\ref{geodesic}) with the metric $g(t)$ of (\ref{FRWv}), in which $c_s^2(t)$ of (\ref{tanh}) is controlled by ${\cal H}$ of (\ref{Hdep}).  To make the situation simpler from the viewpoint of FRW analogue gravity, we follow the authors of \cite{jain1,jain2,jain3} in assuming that the system is essentially two-dimensional (a pancake condensate). This permits us to use the given speed of sound in a 2D setting with a direct correspondence with the FRW metric.

As in \cite{jain1}, on taking $d=2$ in (\ref{FRWv}) we can write
\begin{eqnarray}
 g_{\mu\nu} &=& \bigg(\frac{\rho_0}{c_{BCS}}\bigg)^{2}\left( \begin{array}{cc}
        -c_{BCS}^2        & 0 \\
                  0 & a(t)^2\delta_{ij}.
                  \end{array} \right)
                  \label{FRW}
                  \end{eqnarray}
 If $c_s^2(t) = c_{BCS}^2 B(t)$ then $a^2(t) = B(t)^{-1}$,
 where
\be
B =\frac{N_0^{BCS}}{N_0 + 4{\bar\alpha}^2/{\bar M}^2}.
\ee
$N_0^{BSC}$ is the value of $N_0$ in the deep BCS regime.

\section{Particle (phonon) creation in a uniform field quench}

We shall follow this 2D approximation but it is not crucial for any of the analysis.
 For the simplest uniform quench from the BCS to BEC regimes the time dependence of $B(t)$ as ${\cal H}$ changes is given from (\ref{Hdep}) as
 \be
 B(t) = (1/2)[1 + \tanh (d - t/\tau_Q)].
\ee
 The FRW scale factor $a(t)$ is then
\ba
a^2(t) &=& B(t)^{-1} = 1 + e^{-2d}e^{2t/\tau_Q}.
\label{a(t)}
\ea
We have normalised $a(t)$ to  unity in the deep BCS regime. To be specific we adopt the $^6Li$ parameters that we quoted earlier. Since $d < 0$ then, insofar as the hydrodynamic approximation is reliable,  in the BEC regime ($t > 0$) we have
 \be
 a(t)\approx e^{-d} e^{t/\tau_Q},
 \label{adS}
\ee
corresponding to an effective de Sitter universe with Hubble parameter $H = 1/\tau_Q$.
That is, the simplest experimental situation of a constant quench rate initially looks to give one of the most interesting analogue models!

The standard approach to phonon production in condensates is given in great detail in \cite{jain1}. In driving the system from the BCS to BEC regimes, with our analogue de Sitter expansion, we move from one flat spacetime at time $t_0$ to another at time $t$. The phonon-free initial ground state evolves into a multi-phonon state with amplitudes derived from the Bogoliubov transformations that relate the initial annihilation and creation operators at $t_0$ to their later counterparts.

 We shall not attempt to perform any calculation of phonon production with respect to the metrics above since, in practice, this simple picture is never implemented due to the breakdown of the simple acoustic model with its SoS Lorentz invariance, as we shall see.

\subsection{The breakdown of the simple acoustic model}

The action (\ref{Seff_theta}) is manifestly Galilean invariant, showing the extent to which the speed-of-sound Lorentzian geodesic equations (\ref{geodesic2}) are a long wavelength limit.
The underlying Galilean invariance  manifests itself through  non-linearity in the phonon dispersion relation.

The simplest approximation that goes beyond the acoustic approximation is that of a 'rainbow' metric \cite{jain2,jain3}, in which phonons with different wavenumbers $k$ travel at different speeds $c_k$. In this we ignore the time derivatives of the density field in comparison to space derivatives. In our formalism this is very simple. We set $\eta = 0$ in (\ref{disperse}), which corresponds to approximating $D_F (x - x')$ of (\ref{corr}) by
\begin{eqnarray}
D_F (x - x') = \frac{2}{(2\pi)^3}\delta (t-t')\int {\bf dk}\frac{e^{-i{\bf k}{\bf(x - x')}}}{ \rho_0k^2/2m + {\bar M}^2},
\label{corr2}
\end{eqnarray}
valid for homogeneous systems in which $\rho_0$ and ${\bar M}^2$ depend only on time.

The outcome is the phonon dispersion relation is
\be
\omega^2 = c_k^2 k^2.
\ee
where ($c_s ~\equiv c_{k = 0})$
 \be
  c_k^2 = c_s^2\bigg[ 1 + \frac{k^2}{{\bar M}^2}\frac{4\alpha^2 c_s^2}{{\bar M}^2} + ...\bigg]
  \label{cklin}
 \ee
 The restoration of Galilean invariance is more rapid in the BEC regime since ${\bar M}^2$ becomes vanishingly small in (\ref{corr2}). We find
 \ba
  c_k^2
  &\approx&  c_s^2\bigg[ 1 + \frac{k^2}{K^2}\bigg]
  \label{cklin2}
 \ea
where
 \be
 K\sim \frac{2{\alpha} c_s}{\rho_0/2m}\approx 4m c_s.
 \label{K}
 \ee
This looks more familiar as the Bogoliubov dispersion relation
\be
\omega^2\approx c_s^2k^2 + (k^2/2M)^2.
\label{bog}
\ee
That is, as $c_s$ vanishes, we have
$\omega\approx k^2/2M$,
describing free diatoms with mass $M=2m$.

  To see the effects of this on particle production we repeat the analysis of \cite{jain2,jain3}.
 For modes of wavenumber $k$ the FRW scale $a(t)$ is modified in a mode dependent way from  (\ref{a(t)}) to
 \be
a^2_k(t)\approx [B(t) + k^2/K_0^2]^{-1} \ee where $K_0 \approx
4mc_{BCS} = (4/\sqrt{3}) k_F$  with $k_F$ the Fermi momentum.

In the BEC regime we find
 \be
a^2_k(t) \approx \bigg[e^{2d}e^{-2t/\tau_Q} + \frac{k^2}{K_0^2}
\bigg]^{-1}\, . \label{a-k-approx} \ee
 From our earlier comments, we take $t>0$. Some caution
is required in that the approximation (\ref{a-k-approx}), derived
from (\ref{cklin}), breaks down when the second term is too large.
The difference between our results and those of \cite{jain2,jain3} essentially lies in the prefactor $e^{2d} < 10^{-1}$ for $d\approx -1.4$. That is just sufficient to make sure that Galilean invariance is the dominant feature. The detailed demonstration of this is tedious but we have included it for completeness.

The 'rainbow' Hubble parameters in the BEC regime can be obtained as
\be H_k(t)\equiv \frac{{\dot a}_k(t)}{a_k(t)} \approx
\frac{1}{\tau_Q}\frac{  e^{2d}e^{-2t/\tau_Q}
}{\bigg[e^{2d}e^{-2t/\tau_Q} + \frac{k^2}{K_0^2}\bigg]}. \ee The
transition when the nonlinear dispersion relation becomes important
for wavenumber $k$ happens at time $t_k$, \be \frac{t_k}{\tau_Q} = d
+ \ln{\frac{K_0}{k}},\label{t_k} \ee in terms of which \be H_k(t)
\approx \frac{1}{2 \tau_Q}\frac{e^{-(t-t_k)/\tau_Q}}{\cosh
((t-t_k)/\tau_Q)}. \label{H_k}\ee From our comments above this
approximation breaks down for $t\gtrsim t_k$.

As a guide to phonon production during the sweep from BCS to BEC we
also need the modified dispersion relation \ba && \omega_k(t)
\approx \omega_0 \bigg[e^{2d}e^{-2t/\tau_Q} + \frac{k^2}{K_0^2}
\bigg]^{1/2} \nonumber \\
&\approx & \sqrt{2} \omega_0 \bigg( \frac{k}{K_0} \bigg)
e^{-(t-t_k)/2\tau_Q } \{\cosh ((t-t_k)/\tau_Q)\}^{1/2} ,\nonumber\\
\ea where we have adopted the notation of \cite{jain2,jain3}, in
which $\omega_0 = |k| c_{BCS}$.
There is an infrared bound on $k$. For the exemplary condensates
with $N\approx 10^5$ atoms, their width is $\xi_0 \approx 10^2/k_F$
\cite{rivers3}. This gives \be 0\lesssim\frac{t_k}{\tau_Q} \lesssim
d + \ln 100, \ee or perhaps a little larger. With $d < -1$ this
gives
\be 0\lesssim t_k \lesssim 3 \tau_Q, \ee
at best (without
having to restrict ourselves to the BEC regime {\it a priori}).
Taking the example of Fig.2 with $b\approx 1.2$ this translates into
a transition when the nonlinear dispersion becomes important for a
value of $t$ for which
\be 0\lesssim 1/a_Sk_F \lesssim 2.5, \ee
as we go from the shortest to the longest
wavelengths. We stress that $t_k$ (or the corresponding $1/a_Sk_F$)
marks the boundary between the applicability of Eq.(\ref{geodesic})
with its 'Lorentzian' structure and the restoration of Galilean
invariance.

The relevant quantity is the ratio
\ba &&{\cal R}_k(t) = \frac{\omega_k(t)}{H_k(t)}.
  \label{Rk}
\ea
\\A quantum mode with wavenumber $k$ only experiences significant
amplification (and hence phonon production) when ${\cal R}_k(t) \ll
1$. As before, this approximation breaks down when  $t\gtrsim t_k$.

In the vicinity of $t_k$, where $H_k$ is small, ${\cal R}_k$ is
corresponding large, as it is for $t$ much greater than $t_k$. In
between it achieves its minimum
\be {\cal
R}_k(t^*)\approx \frac{\tau_Q}{\tau_0}\bigg(\frac{k}{k_F}
\bigg)^{2}. \ee
In order to have any phonon production we must have
as fast a quench as possible, with a current lower bound of
$\tau_Q/\tau_0\approx 10$ and a lower bound of $k/k_F$ of $10^{-2}$,
say, for our typical condensate. Then, for the lowest momentum
phonons, \be {\cal R}_k(t^*) \approx 10^{-3}, \ee this minimum
increasing as momentum increases. Thus, from (\ref{Rk}), there is a
window in which ${\cal R}_k(t)$ is sufficiently small to expect
phonon production. However, even then the number is negligible. For such low momentum phonons to be produced within the time
scale $t_k$,  the corresponding number density for
each $k$ mode can be approximated by
\be {\cal N}_k \approx
e^{\int_0^{t_k} H_k (t) dt} < e^{H_k (t^*) t_k} \approx e^{0.6
\big(d+\ln \frac{K_0}{k}\big)} \, ,
\ee
where we have used
(\ref{t_k}) and (\ref{H_k}).  For a lower bound of the
momentum, say $ k=10^{-2} k_F$, again with $ d \approx -1.4$ and $K_0
=(4/\sqrt{3}) k_F$, the corresponding ${\cal{N}}\approx 1$.  For higher momentum
phonons and somewhat slower quenches ${\cal R}_k(t) \gg 1$
throughout and there is {\it no} phonon production. Even when they occur, the effective temperature of the few phonons produced by changing the metric is $O(\varepsilon_F)$ (in units in which $k_B = 1$), much lower than that of the gas. As a result such phonons will be swamped by thermal phonons.

As for reversing the direction of the quench, we have seen that the
non-linear effects are greater as $c_s$ becomes smaller, making a
quench beginning in the deep BEC regime problematical.

Either way, a cold Fermi gas is not a helpful system for exploring parallels with early universe cosmology, if by that we mean that phonon production is determined from the geodesic equation (\ref{geodesic}), despite the optimism of \cite{jain1,jain2,jain3} for general condensates. Simulations for coupled bosonic binary condensates, which also have a gapless and a gapped mode, suggest that they are better candidates for gravitational analogies (\cite{fischer,visser5}), when details are taken into account.

However, instead of requiring the geodesic equation to be relevant, we can ask the general question of whether we can predict and observe phonon production beyond the acoustic approximation.
This may be possible by applying an oscillating external
field, corresponding to the metric of a cyclic universe \cite{jain1}, were the acoustic approximation relevant. This
is an interesting case in that, with the final and initial states
identical, particle production occurs only as a result of parametric
excitation.

A simple choice is to
take \be \frac{1}{k_F a_S} = \frac{1}{k_F a^0_S}(1 - \sin \omega t),
\ee corresponding from (\ref{Hdep}) to an oscillating field in which
the oscillation of $a_S^{-1}$ about $(a^0_S)^{-1}$ extends to the
unitary regime.

As a result
\ba
a^2(t) &=& 1 + a_0^2~e^{-2A \sin\omega t}
\nonumber
\\
&\approx& a_0^2~e^{-2A \sin\omega t},
\ea
where $A = b/k_Fa^0_S$ and $a_0^2 = \exp -2(d -  b/k_Fa^0_S)$.

The corresponding rainbow metric scale factor is
\be
a^2_k(t) \approx \bigg[a_0^{-2}~e^{2A \sin\omega t} + \frac{k^2}{K_0^2}\bigg]^{-1},
\label{a-k-approx2}
\ee
from which the rainbow Hubble parameters follow as
\be
H_k(t)\approx -A\omega\cos\omega t~\frac{\bigg[- a_0^{-2}~e^{2A \sin\omega t} + \frac{k^2}{K_0^2}\bigg] }{\bigg[a_0^{-2}~e^{2A \sin\omega t} + \frac{k^2}{K_0^2}\bigg]}.
\label{Hk2}
\ee
How type of prediction we can make is not yet clear. This is under consideration and we shall not pursue it further here, beyond noting that for our parameters the bulk of any phonon production is controlled again by the Galilean group.

For a completely different test of analogies with cold fermi gases, we observe that a further consequence of this splitting of time derivatives in the action, when it arises, is to induce fluctuations in the time of flight of phonons or sound waves. Whether or not this can be cast in the language of stochastic fluctuations as in \cite{ford,krein} is being analysed at the moment.

\section{Conclusions}

We have continued our exploration of the properties of cold Fermi gases which can be tuned through a narrow Feshbach resonance. We had shown in previous work \cite{rivers3} that the system can be represented by a coupled two-field model of a gapless Goldstone phonon and a gapped Higgs mode.
Here we have gone step further to trace out the Higgs to give an effective purely phononic theory. In particular, the non-linearity of the phonon dispersion relation which can be calculated explicitly.

In choosing to work with tunable condensates, we had hoped  that they would
provide a straightforward system for establishing parallels
with causal properties of the very early universe. In particular we are interested in the spontaneous creation of particle (phonons) because of the rapid variation of the speed-of-sound Lorentzian acoustic metric which defines their geodesics.

The first observation is that, for there to be any measurable effects, the speed of sound must change fast. This happens best in the BEC regime.
However, the parallels with the early universe are complicated, for very different reasons.
For phonon production our ability to construct analogue FRW
universes, for example, is compromised by
the restoration of Galilean invariance at the expense of the
SoS Lorentzian behaviour of (\ref{geodesic}), a counterpart  \cite{visser2,visser3} to the possible (but unobserved)
Lorentzian breakdown of the early universe at the Planck length.
However, whereas the Planck energy is sufficiently high that its
effects can usually be ignored, this restoration of Galilean
invariance  in the BEC regime is sufficiently rapid to make the
predictions from the acoustic metric unreliable.   The effect is that any phonon production has little, if anything, to do with the geodesic equation (\ref{geodesic}) with which we began this article. Nonetheless, if we treat phonon production as an interesting question in its own right, without looking for analogies, we may still be able to make useful predictions.
At the moment these are largely questions of principle since, in our work
we have only considered homogeneous condensates. We should implement
trapping, otherwise the condensate will split apart.

We conclude with a comment on the relation of our work, both here and in our papers \cite{rivers,rivers2,rivers3,rivers4} to that of the condensate mainstream, often canonically Hamiltonian based. Apart from ourselves several authors have used a path integral approach, both without an explicit resonance (e.g. \cite{aitchison,diehl}) and with a resonance (e.g. \cite{ohashi}), but none as far as we know in the same detail when a resonance is present.  A key ingredient in our analysis is our emphasis on step by step Galilean invariance, without which we could not create our fluid models. In fact, the importance of preserving the underlying symmetry in all approximations, which goes back at least to \cite{greiter}, is very clear in discussions of the unitary regime where explicit Galilean invariance is crucial \cite{son,rivers3}. Unfortunately, Galilean invariance is not always manifest in condensed matter papers, particularly from a canonical viewpoint, which makes comparison difficult.
However, whenever we have compared our results to those obtained by other methods,  we are in agreement e.g. the dependence of the sound speed on external field (see \cite{rivers}), the nature of damped oscillations (see \cite{rivers3}), and the strength of condensate interactions (again see \cite{rivers}).

\section*{Acknowledgements}

We thank Dani Steer of APC, University of Paris Diderot, Marek Tylutki of Jagiellon University, Krakov, Piyush Jain of UBC, Vancouver and Silke Weinfurtner of SISSA, Trieste, for helpful discussions. The work of JTH, DSL and CYL was supported in part by the National Science Council, Taiwan.

\section{Appendix}
The explicit fermi density is
$$\rho^F_0 = \int d^3 {\bf p} / (2\pi)^3 \ \left[ 1 -
\varepsilon_{p}/E_{p}
  \right] $$
  where, in conventional notation, $\varepsilon_{k}=k^2/2m$ and $E_{p}=(
\varepsilon_{p}^2 + g^2|\phi_{0}|^2 )^{1/2} $.

$ N_0$ is the density of states at the
Fermi surface, but the other relevant coefficients have no immediate interpretation. After renormalisation \cite{rivers} they take the form
\begin{eqnarray}
N_0 &=& g^2 |\phi_0|^2\int \frac{d^{3} {\bf p}}{(2\pi)^3} \frac{1}{ 2 E_{p}^3} \, ,
\\
\alpha&=& 2|\phi_{0}|
\kappa^{1/2}
\bigg[1 + \frac{1}{2}g^2\int \frac{d^3 {\bf p}}{(2\pi)^3}
\frac{\varepsilon_{p}}{ 2E_{p}^3}\bigg] \, ,
\\
{\bar M}^2 &=& -\frac{g^2 \kappa m}{\pi a_S}-2 g^2 \kappa \int
\frac{d^3 {\bf p}}{ (2\pi)^3} \bigg[\frac{\varepsilon_{p}^2}{
E_{p}^3} - \frac{1}{ ({\bf p}^2/2m)}\bigg]   \, , \\
\eta &=& g^2\kappa \int \frac{d^3 {\bf p}}{(2\pi)^3}
\frac{\varepsilon^2_{p}}{ 2 E_{p}^5} \, , \label{defs}
\end{eqnarray}
and \ba \zeta &=&  \int \frac{d^3 {\bf p}}{
(2\pi)^3}\bigg[\frac{1}{8 E_{p}^3} \bigg[ \bigg( 1-3 \frac{g^2
|\phi_0|^2}{E_{p}^2}\bigg) \frac{\varepsilon_{p}}{m} \nonumber\\
&+& \bigg( 5 \frac{g^2 |\phi_0|^2}{E_{p}^2} \bigg( 1-\frac{g^2
|\phi_0|^2}{E_{p}^2}\bigg) \bigg) \frac{|{\bf p}|^2 \, {\hat {\bf
p}} \cdot {\hat \nabla} }{m^2} \bigg] \bigg] \, . \label{zeta} \ea
In (\ref{zeta}) $ {\hat {\bf p}}$ and  ${\hat \nabla}$ are the unit
vectors along the direction ${\bf p}$ and the direction of the
spatial variation of the phase mode $\theta$ respectively.

In terms of $\zeta$ the scale factor is \be \kappa \;=\;
\frac{\rho_0}{4m g^2\zeta + 2} \, , \ee
In our approximations $\eta$ is not particularly relevant, but sufficient to say that $\eta = N_0$ in the deep BCS regime (when $\alpha$ is small) and falls off in the same way as $N_0$ in the deep BEC regime (when $\alpha\approx\rho_0$).

Finally,
in the definition of $\bar
M^2$, we have used the relationship between the $s$-wave scattering
length $a_S$ and the binding energy,
 \be
 2\mu - \nu = \frac{g^2 m}{4\pi a_S}.
 \label{Hdep0}
\ee
$N_0$ and $\alpha$ depend indirectly on $a_S$ through the constancy of $\rho_0$.
\end{document}